\documentclass[%
  reprint,           % PRL 推荐用 reprint 而不是 preprint
  amsmath,amssymb,
  aps, prl,               % 关键：prl
  superscriptaddress,% 作者地址脚注
  longbibliography,  % 可选：显示完整参考文献标题
  floatfix,
]{revtex4-2}
\usepackage{natbib}
\usepackage{bm}
\usepackage{comment}
\usepackage{color}
\usepackage{bm}
\usepackage{graphicx}
\usepackage{svg}
\usepackage{etoolbox}
\usepackage{newunicodechar}
\usepackage{calc}
\usepackage{amsmath}
\usepackage{booktabs}
\usepackage{mathrsfs}
\usepackage{physics}
\usepackage{enumitem}
\usepackage{hyperref}
\usepackage{xcolor}
\usepackage{ulem}
\begin{document}

%\preprint{APS/123-QED}

\title{\textbf{Seismology of Glasses} }% 

\author{Zhiyu Cao}
\affiliation{Center for Theoretical Biological Physics, Rice University, Houston, TX 77005}

\author{Peter G. Wolynes}
\email{pwolynes@rice.edu}
\affiliation{Center for Theoretical Biological Physics, Rice University, Houston, TX 77005}
\affiliation{Department of Chemistry, Rice University, Houston, TX 77005}
\affiliation{Department of Physics, Rice University, Houston, TX 77005}


\date{\today}% It is always \today, today,

%\tableofcontents

%\section*{Significance}
%During mitosis, chromosomes are compacted into cylindrical rod-like structures that ensure their reliable transfer to daughter cells. Using a motorized chromosome model, we underline the roles in mitosis of two specific molecular machines, that can be modeled as grappling motors, condensin I and condensin II. Particular attention is given to the way the operational mechanisms of these motors lead to the breaking of symmetries in mitosis and how their action gives rise to characteristic defects.

\begin{abstract}
Aging and relaxation of glasses proceed through localized rearrangements that also generate long-ranged mechanical responses in their surroundings. We develop a theory that treats these rearrangements as internal “microseismic” sources, and thereby separates the direct effects emerging from an event from the effects from the far-field displacements it generates. These far-field disturbances propagate and relax through the viscoelastic medium through which they move, producing a stochastic background of weak but numerous perturbations at any given location. Combining the description of local events from the Random First Order Transition theory with this propagating background, we derive a unified equation for the dynamical structure factor of a relaxing glass. The theory identifies a finite wave-vector and time window in which the growing viscoelastic far-field disturbances can produce compressed exponential relaxation, which has been observed using X-ray photon correlation spectroscopy. At higher wave vectors that resolve local jumps, heterogeneity of the local event dynamics instead produces stretched exponential relaxation. The crossover between these behaviors depends on the event statistics and the scattering phase accumulated across the region reached through viscoelastic propagation from the numerous distant sources.

\end{abstract}

\maketitle
%\tableofcontents
%\newpage
% Use \firstpage to indicate which paragraph and line will start the second page and subsequent formatting. In this example, there are a total of 11 paragraphs on the first page, counting the first level heading as a paragraph. The value {12} represents the number of paragraphs starting on the second page. If a paragraph runs over onto the second page, include a bracket with the paragraph line number starting the second page, followed by the paragraph number in curly brackets, e.g. "\firstpage[4]{11}".

% If your first paragraph (i.e. with the \dropcap) contains a list environment (quote, quotation, theorem, definition, enumerate, itemize...), the line after the list may have some extra indentation. If this is the case, add \parshape=0 to the end of the list environment.
\textit{Introduction}---Both aging of glasses and relaxation of supercooled liquids are spatially intermittent \cite{ediger2000spatially,weeks2000three}. At sufficiently low temperatures, structural relaxation proceeds through rare cooperative rearrangements that typically involve tens to hundreds of particles near $T_g$ \cite{adam1965temperature,kirkpatrick1989scaling}. These events are localized in space but generate long-ranged mechanical responses through the emerging rigidity of the glass \cite{eshelby1957determination,picard2004elastic,martens2011connecting,chattoraj2013elastic,ferrero2014relaxation,tahaei2023scaling,berthier2025yielding}. A glass may therefore be viewed as a medium continuously shaken by sparse internal “microseismic” events. Like a seismograph, a tracer particle in the glass records both the effects of nearby rearrangements and the cumulative far-field displacements generated by quite distant events. These seismic effects can be monitored by X-ray photon correlation spectroscopy \cite{ruta2012atomic}, and, in principle, direct imaging \cite{zhang2018spatially,mahato2026anomalous}. 

The Random First-Order Transition (RFOT) theory gives a good account of many relaxation processes as one approaches the glass transition from above as well as below $T_g$, when aging begins \cite{kirkpatrick1987connections,kirkpatrick1987stable,kirkpatrick1989scaling,xia2000fragilities,lubchenko2007theory,lubchenko2004theory}. Near the glass transition, the local rearrangements are reasonably compact (and can be thought of as ``instantons"~\cite{dzero2005activated,dzero2009replica,charbonneau2026rare}), whereas more ramified, string-like or fractal rearrangements~\cite{stevenson2006shapes} associated with secondary relaxation \cite{stevenson2010universal} become important near instabilities as at the crossover temperature~\cite{stevenson2006shapes} or under applied stress~\cite{wisitsorasak2012strength}. Despite their differing geometries and activation statistics, primary and secondary relaxation events both involve a finite amplitude rearrangement in a localized core accompanied by a weaker elastic displacement field outside the core \cite{lancaster1997study,dyre1999solidity}. For many observables such as calorimetry and dielectric relaxation, the elastic effects are only indirectly important. A tracer particle found inside the core of an event therefore undergoes a large, non-perturbative displacement; however, a distant tracer still will experience the much weaker but more numerous mechanical responses transmitted through the surrounding medium. Here, we will explore how rare local rearrangements and the accumulated responses to distant events combine to determine the dynamical scattering and statistics of local displacements of a tracer. This conspiracy explains the compressed-exponential relaxation observed in many glassy systems \cite{cipelletti2000universal,duri2006length,ruta2012atomic}, and how the observed dynamics changes with the probing length scale $q^{-1}$ \cite{duri2006length,caronna2008dynamics,guo2009nanoparticle,bouzid2017elastically,wu2018stretched,ruta2020wave,begam2021kinetics,dallari2023stochastic,shen2026length}?

Although many of the arguments we use here can be applied to systems like gels and the cytoskeleton, here we will focus on thermally activated avalanches in structural glasses, whose statistics are constrained by the RFOT theory. We formulate a dynamical scattering theory for this seismology of glassy systems by explicitly separating local rearrangements from the far-field displacements they generate. Tracer decorrelation depends not only on the geometry and activation statistics of the underlying events, but also on how their mechanical perturbations propagate and ultimately relax. Crucially, these far-field disturbances are not simply carried by pure elasticity but through a viscoelastic glassy medium. Combining a statistical description of local events with the effects of this stochastic, propagating far-field background, we derive a unified integral equation for the dynamical structure factor (DSF) of the tracer. We find that relaxation crosses over with the length scale being probed between two distinct dynamical regimes: at small wave vector and moderate times, the dynamic structure factor is dominated by the collective far-field responses leading to compressed-exponential relaxation, whereas at large $q$, direct local instanton events dominate and the decay becomes more nearly stretched exponential. We focus here on treating the activated events of structural glasses, but collapse-like events invoked for soft gels \cite{cipelletti2000universal,bouchaud2001anomalous} can also be accommodated in a similar framework.

\textit{Event-based scattering description}---We first consider the self-intermediate scattering function of a tagged probe,
\begin{equation}
F_s(q,t)=
\left\langle
\exp\left\{
i\mathbf q\cdot
[\mathbf u(\mathbf r_0,t)-\mathbf u(\mathbf r_0,0)]
\right\}
\right\rangle,
\label{Fqt}
\end{equation}
which is the characteristic function of the tracer displacement distribution. We label each activated event by its class $a$, size $V$ proportional to the number of participating molecules, center $\mathbf R$, orientation tensor $\underline{\underline{Q}}$, and event duration $\tau_{d,a}(V)$, abbreviated as $\tau_d$ when its class and size dependence are not needed. We will argue that in glasses the durations of individual events themselves are relatively short compared to the timescales of interest. The class index specifies the rearrangement geometry and source symmetry, such as compact, string-like, or collapse-like~\cite{stevenson2006shapes}, whereas $V$ sets the size of the local activated events; $\underline{\underline{Q}}$ fixes the angular pattern of the latter. The intermittent sequence of event onsets is denoted by the times $\{\tau_i\}$, with waiting times between instantons $\tau_w=\tau_{i+1}-\tau_i$; the age of the event $i$ at observation time $t$ is $\tau=t-\tau_i$. For a tracer at $\mathbf r_0$, we define the separation between the activated source and the tracer as $\mathbf x=\mathbf r_0-\mathbf R$, with $x=|\mathbf x|$. We will also explore the effect of a finite barrier crossing time $\tau_d$. The event duration $\tau_{d}$ represents an intrinsic timescale of the buildup of an individual activated event. For activated events, this is distinct from both the much longer waiting time between event onsets and the viscoelastic propagation time from source to tracer. The time interval regimes $0<\tau<\tau_{d}$ and $\tau\geq\tau_{d}$ describe a developing and a completed source, respectively. In the simplest RFOT picture, the activated transition path is fast, with $\tau_{d}$ of order the microscopic time $\tau_0$ [see \textcolor{blue}{Supplemental Material (SM)}]. Looking more carefully at the activated transitions, the dynamics of setting up the wetting effect~\cite{villain1985equilibrium,lubchenko2003barrier}, and the stringy nature of the interface~\cite{stevenson2006shapes}, as well as the facilitation effects of neighboring regions~\cite{ritort2003glassy,xia2001microscopic,hasyim2024emergent} may allow event durations to be as long as the so called $\beta$ or secondary relaxation time $\tau_\beta$, timescale but at most a fraction of $\tau_\alpha$. We will show these effects are subdominant. In gels, collapse events may take more time relative to propagation~\cite{bouchaud2001anomalous}.

\begin{figure}[htbp]
  \centering
  \includegraphics[width=0.8\columnwidth]{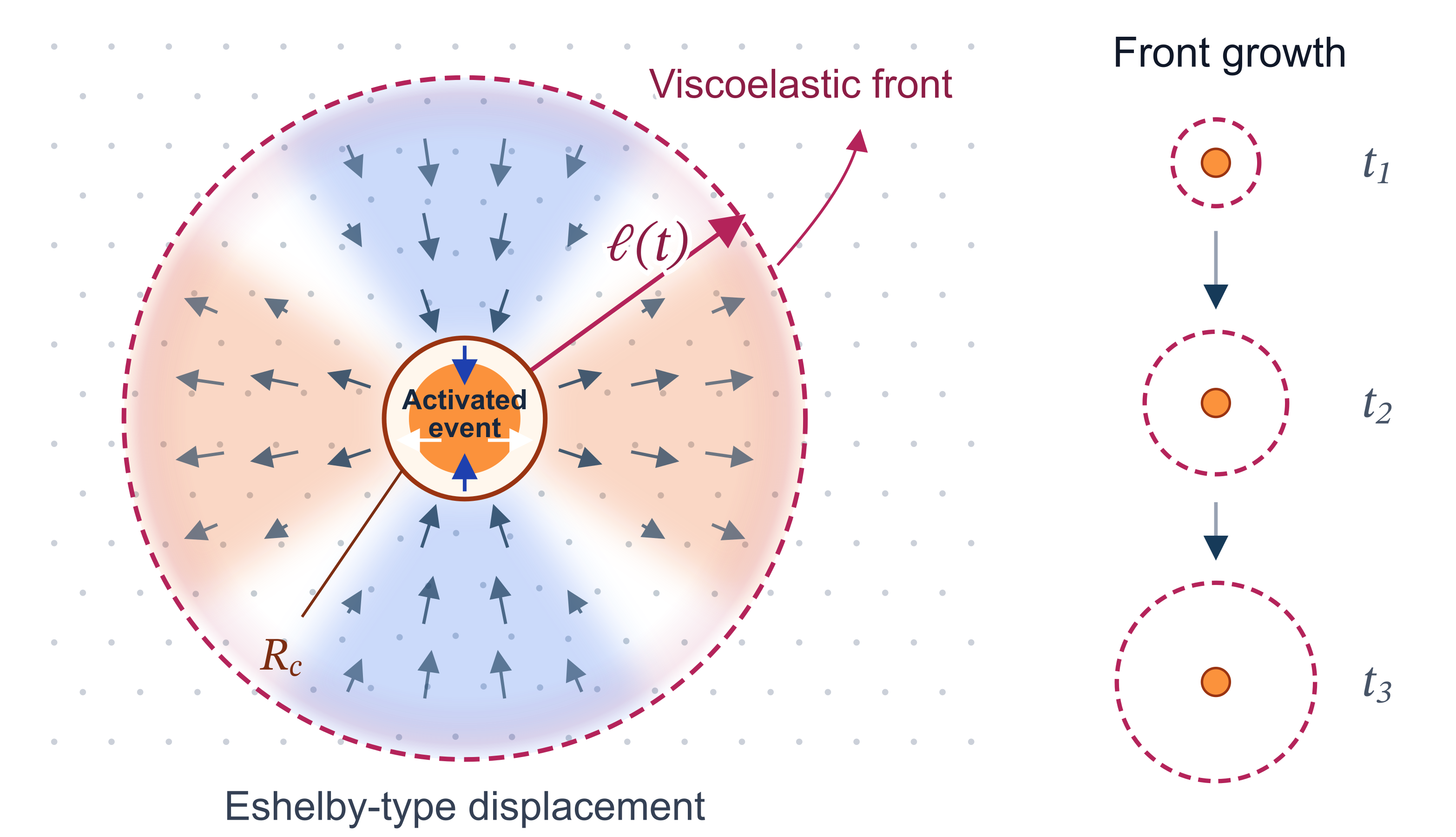}
 \caption{Schematic of the viscoelastic response to an activated rearrangement of radius $R_c$. Arrows and alternating lobes indicate the Eshelby-type displacement field within the front at $\ell(t)$. The right-hand sequence illustrates front growth.}
  \label{fig:1}
\end{figure}

For compact clusters, the effective core radius of the event scales with size; $R_c$ is roughly $V^{1/3}$ \cite{xia2000fragilities,berthier2005direct,biroli2006inhomogeneous,biroli2008thermodynamic}. If the tracer is located inside the rearranging core, it undergoes a finite local displacement of the order of the Lindemann length $d_L$. When $x$ is larger than $R_{c}$, an event is detected only through its weaker retarded displacement of the medium; see Fig. \ref{fig:1}. We write these two conditional single-event factors as
\begin{equation}
\phi_a(q;V,\mathbf x,\underline{\underline{Q}};\tau)
=
\begin{cases}
\displaystyle
\left\langle
e^{i\mathbf q\cdot\Delta\mathbf u_a^{\mathrm{loc}}(V;\tau)}
\right\rangle,
& x<R_{c},\\[6pt]
\displaystyle
e^{i\mathbf q\cdot\mathbf u_a^{\mathrm{far}}(V,\mathbf x,\underline{\underline{Q}};\tau)},
& x>R_{c}.
\end{cases}
\label{singleEventFactor}
\end{equation}
For a completed event, the local displacements within the core,  $\Delta\mathbf u_a^{\mathrm{loc}}$, are taken to be random with variance $\langle\Delta u_{a,\mu}^{\mathrm{loc}}\Delta u_{a,\nu}^{\mathrm{loc}}\rangle_{\zeta}=d_{L}^2\delta_{\mu\nu}/3$ , giving the familiar Debye-Waller factor $\phi_a^{\mathrm{loc}}(q;V)
\simeq e^{-{q^2d_{L}^2}/{6}}$. When $qd_{L}\gtrsim1$, a single direct hit of the local event nearly decorrelates the dynamical structure factor. The differences between compact and string-like rearrangements (as well as collapse sources that occur in gels) enter only through their class-dependent core sizes, durations, and event statistics, all of which can be determined from RFOT theory.

We next average this single-event response over the intermittent sequence of event onsets $\{\tau_i\}$. Near $T_g$, we can assume that fast intrabasin vibrations factor from activated structural motion so that $F_s(q,t)=F_{\mathrm{vib}}(q,t)\mathcal F_s(q,t)$. Here $F_{\mathrm{vib}}(q,t)=\langle\exp\{i\mathbf q\cdot[\delta\mathbf u_{\mathrm{vib}}(t)-\delta\mathbf u_{\mathrm{vib}}(0)]\}\rangle$ is the short-time vibrational correlator, which again depicts a Debye-Waller plateau before structural relaxation, and $\mathcal F_s$ is the normalized structural correlator. We describe activated events by local renewal processes, assuming that distinct source regions are statistically independent and their displacement responses add linearly. Let $W_a(V,\underline{\underline{Q}};\tau_w|\mathbf x)$ be the joint probability density of the next event's class, size, orientation, and waiting time, which can be formulated through the RFOT theory \cite{lubchenko2007theory,ediger2021glass}. The marginal waiting-time density can be written as $\psi_{\mathbf x}(\tau_w)=
\sum_a\int\! dV\!\int\! d\underline{\underline{Q}}\,
W_a(V,\underline{\underline{Q}};\tau_w|\mathbf x)$ and the corresponding renewal memory kernel $M_a$ satisfies $\widetilde M_a
(V,\underline{\underline{Q}};z|\mathbf x)
=z\widetilde W_a(V,\underline{\underline{Q}};z|\mathbf x)/(1-\widetilde\psi_{\mathbf x}(z))$, where $\widetilde \cdot$ denotes the Laplace transformation $\tau_w\to z$. Events occurring at location $\mathbf x$ give rise to independent displacements of the tracer. Combining the independent regional contributions, we finally obtain the normalized structural correlator of the tracer,
\begin{equation}
\mathcal F_s(q,t)
\simeq
\exp\left[
\int d^3x\,n_{\mathrm{reg}}(\mathbf x)
\ln\mathcal F_{\mathbf x}(q,t)
\right],
\label{regionalAssembly}
\end{equation}
where $\mathcal F_{\mathbf x}(q,t)$ is the characteristic factor of the cumulative displacement induced at the tracer by events contributing from the location $\mathbf x$ with $\mathcal F_{\mathbf x}(q,0)=1$ and $n_{\mathrm{reg}}$ is the number density of independent source regions. Considering the effect of both the local events and far-field disturbances based on Eq.~\eqref{singleEventFactor}, the renewal equation of $\mathcal F_{\mathbf x}(q,t)$ can be written as:
\begin{equation}
\begin{aligned}
\partial_t
\mathcal F_{\mathbf x}(q,t)
&=
\sum_a\!\int_0^t \!d\tau_w\!
\int\! dV\!\int\! d\underline{\underline{Q}}\,M_a(V,\underline{\underline{Q}};\tau_w\mid\mathbf x)
\\
&\times
\left[
\phi_a(q;V,\mathbf x,\underline{\underline{Q}};t\!-\!\tau_w)
\!-\!1
\right]\!
\mathcal F_{\mathbf x}(q,t\!-\!\tau_w),
\end{aligned}
\label{renewal}
\end{equation}
For local events $x<R_c$, each direct hit nearly erases the correlation, so that the relaxation essentially is the probability that no rearranging core has directly intersected the tracer during the interval $[0,t]$. For the compact clusters, this survival probability is well represented over the experimentally relevant relaxation window by  a stretched exponential $\mathcal F_{\mathrm{loc}}(t)
\simeq
\exp\left[-(t/\tau_\alpha)^\beta\right]$. Microscopically, the stretched form follows by averaging over the distribution of local rearrangement rates generated by barrier fluctuations and facilitation \cite{xia2001microscopic}. At small $q$, local events are less important because the effect of the extremely numerous and distant far-field disturbances becomes significant. For disturbances with $x>R_c$, we derive the retarded displacement $\mathbf u_a^{\mathrm{far}}(V,\mathbf x,\underline{\underline{Q}};\tau)$ from the causal viscoelastic response to an RFOT rearrangement. The continuum RFOT framework \cite{peter2009spatiotemporal,wisitsorasak2014dynamical,wisitsorasak2017dynamical} describes the glass by a structural mobility field $\mu(\mathbf r,t)\simeq\tau_{\alpha}^{-1}(\mathbf r,t)$ coupled to a fictive-temperature field $T_F(\mathbf r,t)$ by including the activated instanton effects into mode coupling theory \cite{bhattacharyya2005bridging,bhattacharyya2008facilitation}. These fields are coupled to elasticity, treating the instantonic events as localized strain sources $\epsilon_{ij,a}(\mathbf r,t;V,\underline{\underline{Q}})$, leading to the coarse-grained displacement field of the glassy medium $u_i(\mathbf r,t)$ and the internal stress tensor $\sigma_{ij}$ through the force balance in the medium (\textcolor{blue}{SM}). The locally imparted displacement appears point-like at large distances and becomes $\epsilon_{ij,a}(\mathbf k,\omega)\simeq M_a(V)Q_{ij}\gamma_a(\omega)$ in Fourier space, where $M_a(V)$ is the integrated strain amplitude and $\gamma_a(\omega)$ describes the event's growth to finally perturb the medium. For an activated event, $M_a(V)$ is nearly independent of size because the internal rearrangement strains nearly cancel, as pointed out by Lubchenko and Wolynes in their analysis of the coupling of phonons to two-level systems in the cryogenic regime, which underlies the universal thermal conductivity relations seen at low $T$ ~\cite{lubchenko2001intrinsic}.
\begin{comment}
    We can account for the duration timescale of the event itself by taking $\gamma_a(\omega)=-[1-e^{i\omega\tau_d}]/(\tau_d\omega^2)$, corresponding to the linear growth on the timescale $\tau_d$ followed by saturation. In their work~\cite{bouchaud2001anomalous}, Bouchaud and Pitard focused on the duration of collapse events in gels $\tau_d$. For activated events in structural glasses, as we shall see, the effects of finite duration appear subdominant.
\end{comment}

To calculate the viscoelastic response to a rearrangement event, we ignore the fluctuations in mobility, except for the average local response. In the complete absence of dynamical heterogeneity, the averaged complex shear modulus would be $G_{\mathrm{ve}}(\omega)
=G(-i\omega)/(\mu_{\mathrm{eff}}-i\omega)$ in Fourier space, where $G$ is the high-frequency shear modulus with a effective relaxation time  $\mu_{\mathrm{eff}}^{-1}$. The spatial heterogeneity of the RFOT mobility field produces a broad spectrum of local Maxwell relaxation times \cite{wisitsorasak2014dynamical}. Over the frequency window relevant to the scattering experiment, for convenience, we will represent the resulting local viscoelastic response by a fractional Maxwell modulus $G_{\mathrm{ve}}(\omega)=G(-i\omega\mu_{\mathrm{eff}}^{-1})^{\beta_{\mathrm{ve}}}/[1+(-i\omega\mu_{\mathrm{eff}}^{-1})^{\beta_{\mathrm{ve}}}]$, where the stretched exponent $\beta_{\mathrm{ve}}$ characterizes the breadth of the relaxation spectrum and roughly corresponds to a stretched exponential response $\beta_{\mathrm{ve}}<1$ \cite{schiessel1995generalized} (\textcolor{blue}{SM}). The ordinary Maxwell response is recovered for $\beta_{\mathrm{ve}}=1$. RFOT theory predicts $\beta_{\mathrm{ve}}$ from thermodynamics quite well \cite{xia2001microscopic,wisitsorasak2014dynamical}, in agreement with experiments above $T_g$. 

As in pure elastic theory, in three dimensions, a shear-like inclusion generates an anisotropic, quadrupolar displacement field whose leading far-field amplitude decays as $x^{-2}$ \cite{eshelby1957determination}, but with viscous dissipation, the response changes with time \cite{puosi2014time}. By separating the transverse and longitudinal responses to the imposed strain $\epsilon_{ij,a}$ in Fourier space by projection $u_{i,a}^{T}=P_{ij}^{T}u_{j,a}$ with $u_{i,a}^{T}\!=\!u_{i,a}^{T}(\mathbf k,\omega;V,\underline{\underline{Q}})$ the transverse Fourier component of the displacements generated in the medium by one event and $P_{ij}^{T}(\mathbf k)=\delta_{ij}-\frac{k_i k_j}{k^2}$, the transverse sector obeys $[G_{\mathrm{ve}}(\omega)k^2-\rho\omega^2]u_{i,a}^{T}=-2iG_{\mathrm{ve}}(\omega)
P_{i\ell}^{T}(\mathbf k)k_j
\epsilon_{\ell j,a}$. Considering an instantaneous event that would immediately leave a permanent strain, so that the frequency dependence of the strain source becomes $\gamma_a(\omega)=(-i\omega)^{-1}$, we can write (\textcolor{blue}{SM})
\begin{equation}
u_{i,a}^{\mathrm{far}}(V,\mathbf x,\underline{\underline{Q}};\tau)
\simeq
\frac{\mathcal A_a}{x^2}
\mathcal E_{i,a}(\hat{\mathbf x},\underline{\underline{Q}})
H_{\beta_{\mathrm{ve}}}(x,\tau).
\label{farFieldForm}
\end{equation}
Here, $\mathcal E_{i,a}(\hat{\mathbf x},\underline{\underline{Q}})
=
c_{1,a}Q_{ij}\hat x_j
+
c_{2,a}\hat x_i
Q_{mn}\hat x_m\hat x_n $ is the dimensionless angular factor, $\mathcal A_a=M_a/[8\pi(1-\nu)]$,  $c_{1,a}=2(1-2\nu)$ and $c_{2,a}=3$  with $\nu$ the Poisson ratio. The scalar envelope $H_{\beta_{\mathrm{ve}}}(x,\tau)$ approximates the development of the retarded displacement toward its quasistatic Eshelby form as a viscoelastic front leaves the source. This factorization retains the static angular pattern but neglects the distinct radial and temporal functions present in the full tensorial response (\textcolor{blue}{SM}). 
\begin{comment}
    For events existed before the observation, the contribution to the scattering is $\Delta H_{\beta_{\mathrm{ve}}}(x,\tau)=H_{\beta_{\mathrm{ve}}}(x,\tau+t)-H_{\beta_{\mathrm{ve}}}(x,\tau)$.
\end{comment}

The characteristic distance reached by a microseismic disturbance depends on the relaxation properties of the medium,
\begin{equation}
\ell(\tau)
\sim
\begin{cases}
c_T\tau,
&
\tau\ll\mu_{\mathrm{eff}}^{-1},\\[2mm]
c_T\mu_{\mathrm{eff}}^{-\frac12\beta_{\mathrm{ve}}}
\tau^{1-\frac12\beta_{\mathrm{ve}}},
&
\tau\gg\mu_{\mathrm{eff}}^{-1},
\end{cases}
\quad
c_T\equiv\sqrt{\frac{G}{\rho}},
\label{propagationLength}
\end{equation}
where $c_T$ is the high-frequency transverse sound speed, and $\ell(\tau)$ characterizes the spatial spread of the shear response. This length grows ballistically at short times and sub-ballistically after stress relaxation becomes important. In a pure Maxwell medium, $\beta_{\mathrm{ve}}=1$, the long-time growth of the front becomes diffusive $\ell(\tau)\sim(c_T^2\tau/\mu_{\mathrm{eff}})^{1/2}$; the effect can thus be seen as a ``hydrodynamic interaction" familiar in polymer physics. 

To understand the $q$-scaling behavior, the smooth causal envelope can be adequately approximated as $H_{\beta_{\mathrm{ve}}}(x,\tau)
\simeq\Theta[\ell(\tau)-x]$. The scattering phase then vanishes outside the front. When $\ell$ is larger than the event size $R_c$, we need to calculate the single-event far-field scattering integral $I_a(q;V,\tau)\equiv
\int_{x>R_c} d^3x
[
1-\operatorname{Re}
\langle
\phi_a(q;V,\mathbf x,\underline{\underline{Q}};\tau)
\rangle_{\underline{\underline{Q}}}
]
\simeq4\pi\int_{R_{c}}^{\ell(\tau)}x^2dx[1-\langle\cos[\frac{q\mathcal A_a(V)}{x^2}
\hat q_i\mathcal E_{i,a}(\hat{\mathbf x},\underline{\underline{Q}})
]\rangle_{\hat{\mathbf x},\underline{\underline{Q}}}]$. Here, the angular averages assume isotropically distributed source-tracer directions and event orientations (see \textcolor{blue}{End Matter}). For small $q\ll R_c^2/\mathcal A_a$, $I_a(q;V,\tau)\simeq
\frac{2\pi}{45}(5c_{1}^2+4c_{1}c_{2}+2c_{2}^2)({q\mathcal A_a})^2(R_c^{-1}-\ell^{-1})$. On the other hand, for sufficiently large $q$, the angularly averaged phase factor is negligible throughout the viscoelastically displaced region, so each newly reached volume element reached by the disturbance contributes an order-unity phase loss, and thus the $I_a(q;V,\tau)$ grows as $\frac{4\pi}{3}(\ell^3-R_c^3)$. If we write $\ell(\tau)\propto
\tau^{\gamma_{\mathrm{ve}}}$ with $\gamma_{\mathrm{ve}}\equiv
{\partial\ln\ell}/
{\partial\ln t}$, then $I_a(q;V,\tau)$ grows as $\tau^{3\gamma_{\mathrm{ve}}}$. The renewal equations~\eqref{regionalAssembly} and \eqref{renewal} combine the single-event effects with the number of event onsets that may contribute during the observation interval. This can be large because the Eshelby displacement falls off slowly as $1/x^2$, allowing distant sources to contribute greatly; the disturbance from a far-field event is most significant when the front covers it, and it gradually diminishes as the front moves away. When the time dependence of the fictive temperature distribution is not taken into account, the apparent cumulative number density $W_a$ is roughly constant, allowing us to solve Eqs.~\eqref{regionalAssembly} and \eqref{renewal} numerically (\textcolor{blue}{SM}). If phase correlations between spatially separated activated events can be neglected, their contributions to $\mathcal F_s$ are additive, so the far-field decay of the memory kernel obeys
\begin{equation}
\begin{aligned}
\mathcal F^{\mathrm{far}}_s(q,t)
\sim
e^{-C_qt^{\beta_{\mathrm{far}}}}
\end{aligned}
\label{compressedExponent1}
\end{equation}
with an exponent $\beta_{\mathrm{far}}
=3(1-\beta_{\rm ve}/2)$ for $t\gg\mu_{\rm eff}^{-1}$. For an ordinary Maxwell medium $\beta_{\mathrm{ve}}=1$, the exponent $\beta_{\mathrm{far}}=3/2$ is compressed. If $\beta_{\mathrm{ve}}<1$ is stretched, the exponent $\beta_{\rm far}>3/2$ will still be compressed exponentially. The compressed far-field decay is therefore restricted to a mesoscopic wave-vector and time window in which the spatial integral $I_a$ continues to grow with \(\ell(t)\). In the strict \(q\rightarrow0\) limit, the radial integral is analytic in \(q^2\) and approaches a propagation-independent value proportional to \(q^2\), so front expansion alone does not imply a persistent compressed decay. Reduced scattering sensitivity to far-field displacements, together with averaging over heterogeneous rearrangements, may thus weaken propagation-controlled compressed relaxation at small \(q\), qualitatively consistent with the stretched relaxation observed at larger length scales \cite{shen2026length}. At molecular wave vectors satisfying \(qd_L\gtrsim1\), direct intersections with compact rearranging cores become visible, and the local-event-dominated decay approaches the stretched exponential form predicted by RFOT theory \cite{xia2001microscopic}. Between these limits, both local and distant events contribute through Eq.~\eqref{renewal}, so a single KWW fit generally yields a wave-vector- and time-window-dependent effective exponent rather than a universal power law.

\begin{comment}
Taken together, the four panels connect the event-based theory to experimentally accessible observables. The $q$-dependent apparent exponent identifies the crossover between far-field and local relaxation, the XPCS comparison tests its temperature dependence, and the inferred displacement statistics expose the non-Gaussian mechanical background generated by activated rearrangements.    
\end{comment}

\begin{figure}[t]
\centering
\includegraphics[width=\columnwidth]{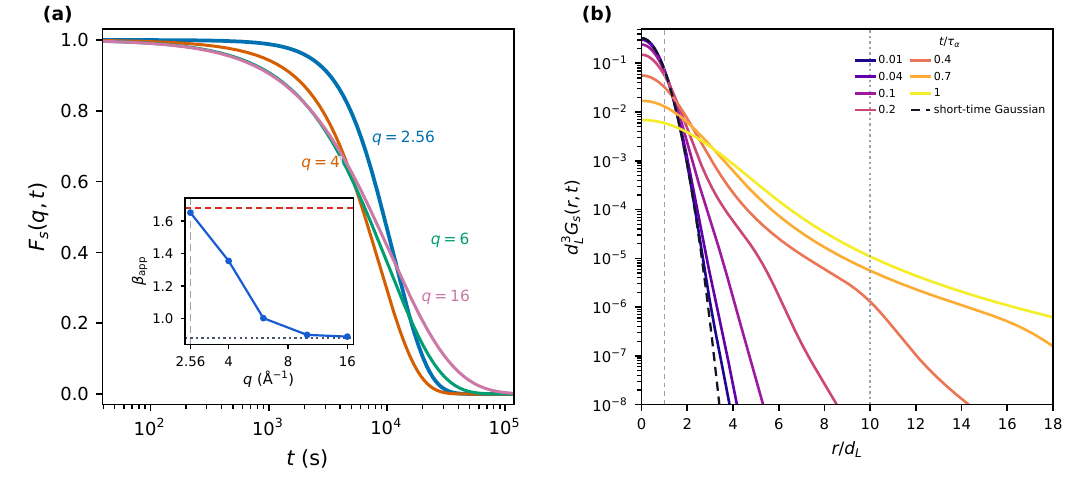}
\caption{\textbf{Scale-dependent relaxation and displacement statistics.}
(a) Calculated structural self-correlators $\mathcal F_s(q,t)$ for four representative wave vectors. The inset shows $\beta_{\mathrm{app}}(q)$, with reference values $1.68$ and $0.88$ and the experimental wave vector $q_0$ indicated.
(b) Evolution of the dimensionless self--Van Hove function $d_L^3G_s(r,t)$ at fixed glassy conditions. The dashed curve is the local Gaussian determined from the central height and curvature at $t/\tau_\alpha=0.01$ and held fixed at all later times. Vertical lines mark $r=d_L$ and $r=2\pi/q_0=10d_L$.}
\label{fig:scale-dependent-relaxation}
\end{figure}

\textit{Numerical results}---The event-based calculation for the metallic glass $\mathrm{Mg}_{65}\mathrm{Cu}_{25}\mathrm{Y}_{10}$ \cite{busch1998thermodynamics} is summarized in Fig.~\ref{fig:scale-dependent-relaxation}. Under the volume-growth scaling assumptions leading to Eq.~\eqref{compressedExponent1}, taking $\beta_{\mathrm{ve}}\simeq0.88$ gives the limiting estimate $\beta_{\mathrm{far}}=3(1-\beta_{\mathrm{ve}}/2)\simeq1.68$~\footnote{This estimate should be distinguished from the effective exponent extracted from the numerical calculation.}. Figure~\ref{fig:scale-dependent-relaxation}(a) shows the calculated structural self-correlator $\mathcal F_s(q,t)$ for representative wave vectors. At the lowest wave-vector shown, $q_0=2.56~\mathrm{\AA}^{-1}$, near the principal peak of the structure factor, individual local jumps produce relatively small scattering phases, and the accumulated far-field response dominates the calculated decay, yielding a fitted exponent \(\beta_{\mathrm{app}}\simeq1.651\). As $q$ increases, finite local jumps become directly resolvable, and the decay crosses toward the stretched exponential form of structural relaxation, predicted for most structural observations, like dielectric relaxation. The inset shows the evolution of $\beta_{\mathrm{app}}(q)$ with wave vector. For comparison with XPCS measurements at $q_0$, the normalized XPCS signal is related to the structural correlator through $[g_2(q_0,t)-1]/{A(q_0,T)}
=
\left|\mathcal F(q_0,t)\right|^2
\simeq
\left|\mathcal F_s(q_0,t)\right|^2$, where $A(q_0,T)$ contains the instrumental contrast and nonergodicity plateau. A direct comparison of the present theory with the measurements by Ruta \textit{et al.}~\cite{ruta2012atomic} is shown in \textcolor{blue}{SM}. The calculation reproduces the crossover from compressed glassy relaxation at $373$-$388~\mathrm{K}$ to stretched liquid-like relaxation at $408$-$410~\mathrm{K}$.  The observed crossover can thus be interpreted as a change in the relative visibility of the collective effects of the far-field rearrangements and the local direct activated rearrangements.

How does a tracer move in real space? We can compute the self Van Hove function through Fourier transformation $G_s(r,t)
\!=\!(2\pi^2)^{-1}\!
\int_0^\infty dq\,q^2F_s(q,t)
\sin(qr)/(qr)$. In Fig.~\ref{fig:scale-dependent-relaxation}(b), we plot $d_L^3G_s(r,t)$ versus $r/d_L$. The dashed curve is a fixed local Gaussian determined from the height and curvature of the central peak at the earliest time, $t/\tau_\alpha=0.01$. As time progresses, a tail emerges. Comparing the later distributions with the short-time reference exposes the emergence and broadening of a non-Gaussian rare-displacement tail \cite{chaudhuri2007universal} extending over several Lindemann lengths. This tail comes from the far-field events. The tail continues to grow but eventually overlaps with diffusion. We show the temperature dependence of the Van Hove function separately in \textcolor{blue}{SM}: the tail is strongest in the far-field-dominated glass and is suppressed as direct local relaxation becomes dominant. Defining the finite-$q_0$ scattering-inferred excess displacement as $\Delta r_G^2=-6\ln|\mathcal F_s(q_0,t)|/q_0^2$ and its logarithmic growth exponent as $\eta_G=d\ln\Delta r_G^2/d\ln t$, we find $\eta_G>1$ in the compressed regime; the inferred displacement therefore grows faster than the diffusive $t^1$ behavior (\textcolor{blue}{SM}). Near to $T_g$, the apparent speed of particles in the tail, $v_{\mathrm{tail}}=d\ell_{\mathrm{vH}}/dt$, remains on the structural scale $\xi/\tau_\alpha$ with $\xi$ the typical compact event radius and is many orders of magnitude slower than the transverse sound speed $c_T$ expected for a hopping particle (\textcolor{blue}{SM}). We see that the wave vector crossover and the real-space tail both provide signatures of the effects of distant rearrangements propagating through a viscoelastic medium.

The results shown here use the rapidly formed-source limit, $\tau_d\ll\tau_\alpha$, in which the far-field response is controlled primarily by viscoelastic propagation. The effect of finite source duration can additionally produce a nonmonotonic $\beta_{\mathrm{app}}(q)$ at intermediate wave vectors while leaving the large-$q$ local limit nearly unchanged (see \textcolor{blue}{End Matter}).

\textit{Discussion}---The seismological perspective of this paper suggests several new experimental and theoretical directions. As it stands, the theory suggests that exploring a wider time window, going into the shorter secondary relaxation timescales, may give insight into the actual internal time course of the activated events, as expected by the RFOT framework. As in geological seismology, having two tracer detectors could allow triangulating the sources. Lubchenko and Silbey highlighted such a ``binocular" approach in their analysis of fluorophores interacting through dipole-dipole interactions with two-level systems in low-temperature glasses \cite{lubchenko2013molecular}. In XPCS mode, analogous information would be obtained by correlating the scattering signals at different wave vectors. Even more directly, imaging of many tracers in a glassy medium, as recently carried out by Kaufman and coworkers \cite{mahato2026anomalous}, could quantify the correlated displacements of distinct localized tracers responding to common activated events. Also, as in geological seismology, it would be helpful to artificially set off perturbations and observe their effects on tracer motion. Lubchenko and Wolynes recently analyzed how photochemical activation by light-induced isomerization of chromophores in glasses greatly changes the distribution of activated events and photofluidizes a glass \cite{lubchenko2020photon}. 

\begin{comment}
    Inspired by two-chromophore spectral-correlation measurements \cite{lubchenko2013molecular}, a two-probe displacement experiment could test the predicted spatial correlations of the far field. Comparing the lag dependence at several probe separations would help distinguish propagation from local source kinetics. An explicit coherent calculation coupling the displacement propagator to the evolving mobility and fictive-temperature fields of continuum RFOT could incorporate aging~\cite{bouchaud1998out,angelani2001topological,lubchenko2004theory,yunker2009irreversible,amir2012relaxations}, elastic heterogeneity~\cite{dmowski2010elastic}, photoinduced fluidization~\cite{lubchenko2020photon}, and mixed compact and string-like rearrangements~\cite{stevenson2006shapes,stevenson2010universal}. The resulting seismic picture is simple: activated rearrangements act as internal sources, viscoelasticity controls the propagation of their mechanical disturbances, and $q$ selects between local and collective far-field dynamics.
\end{comment}

\textit{Acknowledgments}---Z.C. and P.G.W. were supported both by the Bullard-Welch Chair at Rice University, grant C-0016, and by the Center for Theoretical Biological Physics sponsored by NSF grant PHY-2019745.

\textit{Data availability}---There are no publicly available research data or software supporting this manuscript. Requests for further information or data should be sent to the authors.

\bibliographystyle{apsrev4-2}
\bibliography{bibfile}

%There are no publicly available research data or software supporting this manuscript. Requests for further information or data should be sent to the authors.

\appendix
\section{End Matter}
\paragraph{The effect of finite event duration.}
In the main text, we have given explicit results in the rapidly formed-source limit, $\tau_d\ll\tau_\alpha$. In this regime, the far-field response is governed primarily by viscoelastic propagation. We believe this limit is appropriate for glasses because, in the RFOT picture, the activated events arise from a balance of a driving force of the configurational entropy and a restraining mismatch energy \cite{xia2000fragilities}, so the barrier crossing time itself for activated rearrangements is still of the order of the microscopic time, making the instantaneous-source approximation natural. The renormalization group argument of Kirkpatrick, Thirumalai and Wolynes \cite{kirkpatrick1989scaling} and the ``fuzzy sphere" analysis of Stevenson, Schmalian and Wolynes \cite{stevenson2006shapes} suggest that establishing the mismatch energy itself occurs on the $\beta$-relaxation timescale. At an extreme, one might imagine that facilitation by neighboring events would allow some event durations to be a fraction of $\tau_\alpha$. To account for finite duration, we can simply convolve the response from instantaneous events with a source that grows linearly over $\tau_d$ before saturating. Using $\tau_d/\tau_\alpha=0$, $0.01$, $0.1$, and $0.3$, for illustration, we see that finite duration source buildup further leads to a nonmonotonic variation of $\beta_{\mathrm{app}}(q)$ with $q$ (\textcolor{blue}{SM}). Recent coherent XPCS measurements do report a nonmonotonic scale dependence, with compressed relaxation near medium-range order and stretched relaxation at larger length scales~\cite{shen2026length}. Such nonmonotonic behavior at shorter length scales can also be traced back to the combination of mode coupling and activated events \cite{bhattacharyya2010subquadratic}. At small $q$, the effects are small. Increasing the duration produces a maximum in $\beta_{\mathrm{app}}(q)$ at intermediate wave vectors, while the large-$q$ local limit remains nearly unchanged. We see that the observed scattering relaxation may well contain distinct contributions from both the finite buildup of a local source and from its subsequent propagation through the surrounding viscoelastic medium, but propagation effects dominate in structural glasses. While a $\beta$-relaxation-scale source duration could provide a modest additional contribution, taking a duration as long as $\tau_\alpha$ would make ambiguous the idea that events are discrete. We see, however, that such a long duration would produce a substantially larger compression exponent than is observed experimentally. A finite source time for the buildup can generate a similar wave-vector dependence as propagation, but only for slowly developing events such as microcollapses in soft gels~\cite{bouchaud2001anomalous}. In general, the static structural correlations from the atomic distribution must also be convolved with the tracer behavior described in this paper to give the detailed $q$-dependence for specific glasses.

\paragraph{Analytical reduction of the far-field scattering integral.}
To find the average far-field scattering, we diagonalize the $\underline{\underline{Q}}$ tensor of an event as $\underline{\underline{Q}}=R\underline{\underline{Q_0}}R^{\mathsf T}$, where $\underline{\underline{Q_0}}$ specifies the principal values of the event tensor and $R$ is distributed according to the normalized uniform measure on rotations. In the corresponding principal-axis coordinates, the source-tracer direction is $\hat{\mathbf x}
=(\sin\theta\cos\varphi,\sin\theta\sin\varphi,\cos\theta)$. For a fixed relative direction $\hat{\mathbf x}$, an overall rotation changes the direction of $\boldsymbol{\mathcal E}_a$ but leaves its magnitude $h_a=|\boldsymbol{\mathcal E}_a|$ unchanged, so
\begin{equation}
\begin{split}
h_a^2(\theta,\varphi)
&=
c_{1,a}^{\,2}
\hat{\mathbf x}^{\mathsf T}\underline{\underline{Q_0}}^2\hat{\mathbf x}\\
&+
(2c_{1,a}c_{2,a}+c_{2,a}^{\,2})
\left(
\hat{\mathbf x}^{\mathsf T}\underline{\underline{Q_0}}\hat{\mathbf x}
\right)^2.
\end{split}
\label{eq:app-angular-magnitude}
\end{equation}
The angular average therefore becomes
\begin{equation}
\begin{split}
&\left\langle
\cos\!\left(
\tfrac{q\mathcal A_a}{x^2}\hat q_i\mathcal E_{i,a}
(\hat{\mathbf x},\underline{\underline Q})
\right)
\right\rangle_{\hat{\mathbf x},\underline{\underline{Q}}}\\
&\quad=
\frac{1}{4\pi}
\int_0^\pi d\theta\,\sin\theta
\int_0^{2\pi}d\varphi\,
\operatorname{sinc}\!\left(
\tfrac{q\mathcal A_a}{x^2}
h_a(\theta,\varphi)
\right),
\end{split}
\label{eq:app-overall-rotation}
\end{equation}
with $\operatorname{sinc}z=\sin z/z$. For simplicity, we first assume the glass is incompressible so that the Poisson ratio \(\nu=1/2\). With the coefficient convention used here, \(c_{1,a}=0\) and \(c_{2,a}=3\). The leading far-field Eshelby displacement is purely radial, $\mathcal E_{i,a}
=3\hat x_iQ_{mn}\hat x_m\hat x_n$. For isotropically oriented pure-shear events with \(\underline{\underline{Q_0}}=\operatorname{diag}(1,-1,0)/\sqrt2\), the projection onto the scattering direction can be found analytically,
\[
\left\langle
\cos\!\left[\tfrac{q\mathcal A_a}{x^2}\hat q_i\mathcal E_{i,a}
\right]
\right\rangle_{\hat{\mathbf x},\underline{\underline{Q}}}
={}_2F_3\!\left(
\frac12,\frac12;
\frac34,\frac54,\frac32;
-\frac{9q^2\mathcal A_a^2}{8x^4}
\right),
\]
where \({}_2F_3\) is a generalized hypergeometric function, a special function defined by a power series with two numerator and three denominator parameters. Here, it provides a compact, closed-form expression for the isotropic angular average. For other values of the Poisson ratio, the angular averaging is more involved (\textcolor{blue}{SM}).

\end{document}